
\magnification=1200
\def\setup{\count90=0 \count80=0 \count91=0 \count85=0
\countdef\refno=80
\countdef\secno=85}
\def\R{\vrule height5.85pt depth.2pt \kern-.05pt \tt R}

\def\G{{\cal G}}
\def\Box#1{\vcenter{\hrule
                    \hbox{\vrule height#1pt \kern#1pt \vrule height #1pt}
                    \hrule}}
\def\lo{\raise2pt\hbox{$<$}\kern-7pt\raise-2pt\hbox{$\sim$}}
\def\go{\raise2pt\hbox{$>$}\kern-7pt\raise-2pt\hbox{$\sim$}}
\def\parallel#1#2{\hbox{\kern1pt \vrule height#1pt
                                 \kern#2pt
                                 \vrule height#1pt \kern1pt}}
\def\v{\varphi}
\def\ol{\overline}
\def\vline{{\vrule height8pt depth4pt}\; }

\def\sub#1{{\lower 8pt \hbox{$#1$}}}
\def\Basis{\hat\Phi}

\def\Del{{\raise.5ex\hbox{$\bigtriangledown$}}}
\def\DEL#1{{\raise.5ex\hbox{$\bigtriangledown$}\raise 8pt \hbox{\kern -10pt
                 \hbox{$#1$}} }}
\def\autoeq{ {\global\advance\count90 by1} \eqno(\the\count90) }
\def\autoeql{ {\global\advance\count90 by1} & (\the\count90) }
\def\ceist{ {\global\advance\count91 by1} (\the\count91) }
\def\autosec{ {\global\advance\secno by 1} (\the\secno) }

\def\Lie#1{{\cal L}{\kern -6pt
            \hbox{\raise 1pt\hbox{-}}\kern 1pt} _{\vec{#1}}}

\def\Z{Z \kern-5pt \hbox{\raise 1pt\hbox{-}}\kern 1pt}


\def\autoref{ {\global\advance\refno by 1} \kern -5pt [\the\refno]\kern 2pt}

\def\readref#1{{\count100=0 \openin1=refs\loop\ifnum\count100
<#1 \advance\count100 by1 \global\read1 to \title \global\read1 to \author
\global\read1 to \pub \repeat\closein1  }}

\def\reftitle{{ \kern -3pt \vtop{ \hbox{\title} \hbox{\author\ \pub} } }}
\def\ref{{ \kern -3pt\author\ \pub \kern -3.5pt }}

\def\refanon{{ \hbox{\pub}\kern -3.5pt }}

\def\circum#1{{ \kern -3.5pt $\hat{\hbox{#1}}$ \kern -3.5pt}}

\setup
\centerline{  }
\line{\hfill \hbox{ITP-UH-04/93} }
\line{\hfill \hbox{March 1993} }
\vskip 3cm
\centerline{\bf Potential Flow Of The Renormalisation Group}
\centerline{{\bf  In A Simple Two Component Model}
\footnote*{Work carried out with an  Alexander von Humboldt
research stipendium.}}
\vskip 1.5cm
\centerline{Brian P. Dolan}
\vskip .8cm
\centerline{\kern 10pt {\it Institut f\"ur Theoretische Physik}
\footnote{**}{On leave of absence from: Department of Mathematical
Physics, St. Patrick\rq s
College, Maynooth, Ireland}}
\centerline{\it Universit\"at Hannover}
\centerline{\it Hannover, Germany}
\centerline{e-mail: dolan@kastor.itp.uni-hannover.de}
\vskip 2cm
\centerline{ABSTRACT}
The renormalisation group (RG) flow on the space of couplings
of a simple model with two couplings is examined.
The model considered is that of a single component
scalar field with $\v^4$ self interaction
coupled,
via Yukawa coupling,
to a fermion in flat four dimensional space.
The RG flow on the two dimensional space of couplings, $\G$,
is shown to be derivable from a potential to sixth order in the
couplings, which requires two loop calculations of the
$\beta$-functions. The identification of a potential requires
the introduction of a metric on $\G$ and it is shown that
the metric defined by Zamalodchikov,
in terms of two point correlation functions
of composite operators, gives potential flow to this order.

\vskip .5cm
\noindent
\vskip .5cm \noindent
\vfill\eject

\noindent\bf\S 1
The $c$-theorem And Potential Flow Of The Renormalisation Group\rm
\vskip .5cm

The question of the nature of the renormalisation group (RG)
flow on the space
of coupling constants for a quantum field theory is a recurrent
one in physics. It has been shown\autoref\newcount\Zam\Zam=\refno
in two dimensional Euclidean
field theory, assuming certain positivity conditions on the
Hilbert space of the theory,
that there exists a  function
on the space of coupling constants which is non-increasing
along the RG trajectories
(the $c$-theorem). This has very important and far reaching
implications for the theory because it puts constraints
on the way that the RG flow can be realised, for example it
can never come back to visit a point where it has already been,
thus eliminating the possibility of limit cycles.
The non-increasing function, $c$,
can be interpreted as a measure of the number of degrees of freedom
of the theory and its decreasing nature as the length scale, $l$, is
increased as being due to \lq\lq integrating out" the degrees of freedom
on scales less than $l$. At fixed points (conformal field theories)
it is the central charge of the theory.

A stronger condition than the $c$-theorem is that of potential
flow.
The possibility of potential flow was emphasised
in\autoref\newcount\Wallace\Wallace=\refno
and\autoref\newcount\Wallacethree\Wallacethree=\refno.
In the latter reference the three loop $\beta$-functions
for massless $\varphi^4$ theory, with several $\varphi^4$ couplings,
were shown to be derivable from a potential,
and it has been conjectured that this property
should hold to all orders in perturbation theory
\autoref\newcount\Banks\Banks=\refno,\autoref\newcount\deSan\deSan=\refno.
An attempt at deriving
general integrability conditions on the $\beta$-functions
for potential flow is presented in \autoref\newcount\pf\pf=\refno.

If the space of couplings is equipped with an invertible
positive definite metric, $G_{ab}$, and
the $\beta$ functions are derivable from a potential, $V(g)$,
in coupling constant space,
\newcount\pot
$$\beta^b G_{ba}:=\beta_a ={\partial\over\partial g^a}V(g), \autoeq $$
\pot=\count90
then a $c$-theorem follows easily.
(Here $a=1,\ldots,n$ labels the dimensionless real couplings $g^a$
which can be thought of as co-ordinates on the $n$-dimensional space of
interactions, denoted
by $\G$.) The metric is necessary since the
$\beta$-functions are naturally defined as vectors,
\newcount\betav
$$\beta^a=\kappa{d g^a\over d\kappa },\autoeq$$
\betav=\count90
and the gradient of a function is a co-vector.
The $c$-theorem follows from (\the\pot) by differentiation of the
potential,
\newcount\grad
$$\kappa{dV\over d\kappa} = \beta^a\partial_a V = G_{ab}\beta^a\beta^b\ge 0,
\autoeq$$
\grad=\count90
where the last inequality holds when positive definiteness
of the metric is assumed. Hence the potential, $V(g)$, is a
function on $\G$ which is non-decreasing along the RG flow and so
$c(g)=-V(g)$ is a function on $\G$ which is non-increasing
along RG trajectories.
Of course this analysis depends crucially
on the choice of metric and various possibilities
for a metric tensor on the space of interactions have been
considered in the literature, [\the\Zam] [\the\Wallace] [\the\Wallacethree]
\autoref\newcount\Metric\Metric=\refno
\autoref\newcount\Denjoe\Denjoe=\refno
\autoref\newcount\Zamo\Zamo=\refno
\autoref\newcount\HughIan\HughIan=\refno.
In particular reference [\the\Wallacethree] examines a concrete
model,
however the technical aspects are necessarily rather
complicated, since they involve the three loop calculations
of \autoref\newcount\Zinn\Zinn=\refno.
At the two loop level there is no question
about whether or not the RG flow in this model is potential,
[\the\Wallace], since a quick
look at the possible Feynman diagrams contributing to the
$\beta$-functions easily shows, without any calculation whatsoever,
that it is always possible to find a potential simply by
adjusting the co-efficients of the appropriate vaccuum diagrams
which can contribute to a potential.
This is because every diagram that contributes to the
four point function, and therefore to the RG flow,
at the two loop level can be obtained by removing a vertex
(differentiating with respect to a coupling) from a vacuum
diagram and each possible vacuum diagram contributes only one
diagram to the RG flow. Hence whatever the numerical contribution
of any given diagram to the two loop four point function, this can always
be obtained from a potential involving vacuum diagrams just by
adjusting the co-efficient of the relevant vacuum diagram appropriately,
i.e.~there is a one to one correspondence between the diagrams
contributing to the four point function at the two loop level and
the vacuum diagrams from which these can be obtained by differentiation
with respect to the couplings.
At the three loop level this is no longer true, one finds more than
one diagram contributing to the three loop
four point functions coming from a single vacuum diagram and there is no
guarantee that the coefficients of the four point diagrams
are compatible with the possibility of being obtained by
differentiation of a single
vacuum diagram. The observation of Wallace and Zia was that it
is possible to {\it choose} a metric which makes the co-efficients
match
consistently, at the order in which they were working.

The purpose of this paper
is to present a simpler model where, even at two loops, looking
at the Feynman diagrams is not sufficient and their contributions
to the $\beta$-functions must be evaluated to determine whether
or not the flow is potential.
The model studied is that of a  scalar field
with $\varphi^4$ interaction coupled with Yukawa couplings
to a single fermion in four dimensions. Thus the space of couplings
is two dimensional. This model was briefely mentioned
at the end of reference [\the\Wallacethree], but no calculations
were presented and so no conclusions drawn.
We shall see that the technical aspects
are simpler than the three loop calculations of the model
studied in [\the\Wallacethree]. In particular it allows for a physical
interpretation of the metric which is exactly that of
Zamolodchikov's metric [\the\Zam], [\the\Zamo]. The approach
presented here will be different to that of [\the\Wallacethree],
in that we shall not assume potential flow to derive a metric but rather
assume Zamolodchikov's metric and prove that this leads to
potential flow of the renormalisation group equations,
at least to fifth order in the couplings appearing in the
$\beta$-functions (sixth order in the potential).

\vskip 1cm

\S 2 \bf The Model And The Metric\rm
\vskip .5cm

Consider a single massless scalar field coupled, via Yukawa
couplings, to a massless four component fermion in four dimensional
Euclidean space. The action is
$$S=\int d^4 xH(x)=\int d^4 x\left(
{1\over 2}\partial_\mu\v\partial^\mu\v+{\lambda\over 4!}\v^4
+i\ol\psi\gamma^\mu\partial_\mu\psi +g\ol\psi\psi\v\right),\autoeq$$
where $\lambda$ and $g$ are real couplings. We shall take
$\lambda\sim g^2$ and calculate to order $g^5$.
Following Zamolodchikov [\the\Zam],~[\the\Zamo] we shall define
a metric on the two dimensional space of couplings
\newcount\zammet
$$G_{ab}=
\vert x\vert^{8}<\Basis_a(x)\Basis_b(0)>\vline\sub {\vert x\vert =l}\;,
\autoeq$$
\zammet=\count90
where $\Basis_a(x)=\partial_aH(x)-<\partial_aH(x)>$ and $l=\kappa^{-1}$ is
a renormalisation length. The metric is defined in terms of
renormalised composite operators and the differentiations
are with respect to renormalised couplings which
are dimensionless.
The factor of $\vert x\vert^{8}$ is included to
make the metric dimensionless.

To lowest order in the couplings the metric can be evaluated
by taking $\Basis_a$ to be normal ordered and
using Wick's theorem.
Thus
$$\eqalign{G_{\lambda\lambda}&={\vert x\vert^{8}\over (4!)^2}
<:\v^4(x)::\v^4(0):>
={1\over 4!}{1\over (4\pi^2)^4}\;+\;o(\lambda) \cr
G_{gg}&=\vert x\vert^8<:\ol\psi\psi\v(x)::\ol\psi\psi\v(0):>
={1\over 4\pi^6}\;+\;o(g^2)\cr
G_{g\lambda}&={\vert x\vert^8\over 4!}<:\ol\psi\psi\v(x)::\v^4(0):>=0
\;+\;o(\lambda g,g^3),
\cr}\autoeq$$
independent of the value of $\vert x\vert$.
The propagators are defined with the standard conventions,
$$<\v(x)\v(0)>={1\over 4\pi^2\vert x\vert^2}\qquad\qquad
\hbox{and} \qquad\qquad<\psi(x)\ol\psi(0)>={i\over 2\pi^2}
{\gamma^\mu x_\mu\over\vert x\vert^4}.\autoeq$$
The interactions contribute terms of order $g^2$
but we will not need to calculate them.
Thus
\newcount\metric
$$G_{ab}={1\over (16\pi^2)\pi^4}\left(
\matrix{4&0\cr 0&{1\over 4!(16\pi^2)}\cr}
\right)+o(g^2)\autoeq$$
\metric=\count90
so this metric is flat and, one might have thought, trivial
but we shall see that the factors in (\the\metric) are crucial
for the property of potential flow.
This metric could also have been obtained using momentum space
renormalisation conditions in (\the\zammet) rather then $x$-space
conditions. For massless theories the result is the same, up to
overall factors of $4\pi^2$.
The lowest order form of the metric (\the\metric) also
co-incides with that appearing in [\the\HughIan], but this equality
will
not hold at higher orders.

It will be convenient in the next section to get rid of some of the
factors of $16\pi^2$ in metric. To this end define
\newcount\coord
$$\tilde\lambda={\lambda\over 16\pi^2}\qquad\qquad
\tilde g ={g\over\sqrt{16\pi^2}}\;.\autoeq$$
\coord=\count90
In these co-ordinates the metric (\the\metric) becomes
\newcount\tmet
$$G_{\tilde a\tilde b}={1\over\pi^4}\left(
\matrix{4&0\cr 0&{1\over 4!}\cr}
\right)+\;o(\tilde g^2).\autoeq$$
\tmet=\count90

\vskip 1cm
\vfill\eject
\S 3 \bf The $\beta$-functions To Order $g^5$ And Potential Flow \rm
\vskip .5cm

At one loop the diagrams contributing to the $\beta$-functions
are given in Figures $1$ and $2$ for $g$ and $\lambda$ respectively.
$\beta^g$ is of order $g^3$ and $\beta^\lambda$ is already of order $g^4$
(remember $\lambda\sim g^2$).
The contributions to the $\beta$-functions are easily calculated
using standard techniques (dimensional regularisation and minimal
subtraction were used
for the results presented here),
$$\eqalign{
\beta^g&={5g^3\over 16\pi^2}\; +\; o(g^5)\cr
\beta^\lambda&={1\over 16\pi^2}\bigl(3\lambda^2 + 8 \lambda g^2 - 48g^4\bigr)
\;+\; o(g^6).\cr}\autoeq$$

To have a consistent expansion in powers of the coupling we must
calculate the next highest contribution to $\beta^g$, which involves
the two loop diagrams in Figure 3 as well as some other diagrams
contributing a term in $g^5$ which we do not need to calculate.
Note that a two loop contribution
to $\beta^\lambda$ would be of order $g^6$, thus a loop expansion
is not the same as an expansion in powers of the coupling.
Diagram (i) in Figure 3 has an overlapping divergence, but the
calculation is standard. Diagram (ii) is easier, it does not have
any divergent subdiagrams. Including these two diagrams
in $\beta^g$ gives the
following $\beta$-functions, consistent to order $g^5$,
$$\eqalign{
\beta^g&={1\over 16\pi^2}\left[5g^3
+{1\over 16\pi^2}\Bigl({\lambda^2 g\over 12} - 2\lambda g^3
+ag^5\Bigr)\right]
\;+\;o(g^7)\cr
\beta^\lambda&={1\over 16\pi^2}\bigl(3\lambda^2 + 8 \lambda g^2 - 48g^4\bigr)
\;+\;o(g^6),\cr}\autoeq$$
or, in terms of the co-ordinates (\the\coord),
\newcount\betf
$$\eqalign{
\beta^{\tilde g}&=5\tilde g^3 +
{1\over 12}\tilde\lambda^2 \tilde g - 2\tilde\lambda \tilde g^3
+ a\tilde g^5\;+\;o(\tilde g^7)\cr
\beta^{\tilde\lambda}
&=3\tilde\lambda^2 + 8 \tilde\lambda \tilde g^2 - 48\tilde g^4
\;+\;o(\tilde g^6).\cr}\autoeq$$
\betf=\count90
The value of the constant $a$ does not affect the ensuing analysis.

The vacuum diagrams from which figures 1-3 can be obtained, by
differentiating with respect to the couplings, are shown in
figure 4. Diagrams (i) of figure 2 and diagram (i) of
figure 3 both come from the same
diagram (v) of figure 4
and there is no reason why their
numerical contributions to the $\beta$-functions should be
compatible with them both coming from a single term in a potential.
This is reflected in equations (\the\betf) where
$\beta^{\tilde\lambda}$ would require a term
$4\tilde\lambda^2\tilde g^2$
in a potential whereas  $\beta^{\tilde g}$
would require a term ${1\over 24}\tilde\lambda^2\tilde g^2$ - with
different coefficients.
Similar coments apply to diagrams (iii) of figure
2 and diagram (ii) of figure 3, both of which come from
diagram (vi) of figure: $\beta^{\tilde\lambda}$ would require a term
$-48\tilde\lambda \tilde g^4$ in a potential
whereas $\beta^{\tilde g}$ would require
$-{1\over 2}\tilde\lambda\tilde g^4$ - again with different
coefficients.

However (\the\betf) are vectors, not co-vectors, and it is only co-vectors
that can be obtained from differentiation of a potential.
So we use the metric (\the\tmet) to convert (\the\betf) into
co-vectors, giving
$$\left(\matrix{
\beta_{\tilde g}\cr
\beta_{\tilde \lambda}\cr}\right)
={1\over \pi^4}\left(\matrix{4&0\cr0&{1\over 4!}\cr}\right)
\left(\matrix{
\beta^{\tilde g}\cr
\beta^{\tilde \lambda}
\cr}
\right)
={1\over \pi^4}\left(\matrix{
20\tilde g^3 +
{1\over 3}\tilde\lambda^2 \tilde g - 8\tilde\lambda \tilde g^3
+a^\prime\tilde g^5\cr
{1\over 8}\tilde\lambda^2 + {1\over 3} \tilde\lambda \tilde g^2 - 2\tilde g^4
\cr}
\right),
\autoeq$$
where $a^\prime$ is a constant which is the sum of the co-efficient of
$\tilde g^2$ in $G_{\tilde g\tilde g}$ and $4a$.
It is immediately obvious that now
$$\beta_{\tilde a}=\partial_{\tilde a}V(g)\autoeq$$
with the potential given by
\newcount\potential
$$V(g)=C + {1\over \pi^4}\left(5\tilde g^4 + {1\over 4!}\tilde\lambda^3
+ {1\over 6}\tilde\lambda^2\tilde g^2 - 2\tilde\lambda\tilde g^4
+{a^\prime\over 6}\tilde g^6\right)
+\; o(\tilde g^8)\autoeq$$
\potential=\count90
(there are no terms of $o(\tilde g^7)$,  the potential can
only contain even powers of $\tilde g$).\hfill\break
Here $C$ is an undetermined integration constant, independent of
the couplings. When $\tilde\lambda=\tilde g=0$, $C$ might be interpreted
as the central charge for a free massless scalar and a free massless
fermion (with both left and right components) in four dimensions,
but obviously it cannot be calculated by the methods used here.

It is remarkable that such a simple metric, with such a strong
physical motivation, [\the\Zamo], does the trick. It would be
very interesting to see whether or not this property holds to
next order in the couplings, but this would require three loop
calculations of the $\beta$-functions as well as the calculation
of the higher order terms in the metric, up to $o(\tilde g^4)$,
and this will not
be attempted here.

\vskip 1cm

\S 4 \bf Conclusions \rm
\vskip .5cm
It has been shown that, to sixth order in $\tilde g$,
the $\beta$-functions
of the model presented in \S 2 can be obtained from the potential
(\the\potential), with the constant $C$ undetermined
($a^\prime$ can be calculated straightforwardly, but
we have not done so because it is not necessary).
It is tempting to speculate that this property may continue
to higher, indeed all, orders in perturbation theory.
If this were true it would be very interesting to determine
what class of field theories has this property. If it is true, it is
unlikely to be a fluke of one or two models and it is more
probable that it has a deeper significance. The set of such models
would clearly constitute a very important class of field theories.

An obvious
question is, would it hold for massive theories?
It is argued in [\the\Wallacethree] that the very existence
of mass independent regularisation schemes (such as
dimensional regularisation) shows that the
introduction of masses should not affect the conclusions,
but the details would become more complicated
and it is not clear (at least to the author) that the simple
interpretation of the metric as the  two point function
presented in (\the\zammet) would continue to hold in the massive
case. Perhaps a detailed analysis would show that it does still
hold,
or perhaps this form of the metric would have to be changed
in the massive case and a different metric, e.g.~that of O'Connor
and Stephens [\the\Denjoe], would
be required for potential flow.

The ideas presented here are not new, but the model is a little
simpler and, I believe, the physics a little clearer than in
the three loop $\lambda\v^4$ model of Wallace and Zia
and a connection has been made with Zamolodchikov's metric.

It is a pleasure to thank Denjoe O'Connor for many useful
discussions concerning the renormalisation group and for
bringing my attention to the work of Wallace and Zia.

\vfill\eject

\bf References \rm\hfill
\vskip1cm
\frenchspacing
\item{[\the\Zam]} A.B. Zamolodchikov, Pis'ma Zh. Eksp. Teor.
Fiz.\bf 43\rm, (1986) 565\hfill
\break  (JETP Lett. \bf 43\rm, (1986), 730)
\smallskip
\item{[\the\Wallace]} D.J. Wallace and R.K.P. Zia, Phys. Lett. \bf 48A\rm,
(1974), 325
\smallskip
\item{[\the\Wallacethree]} D.J. Wallace and R.K.P. Zia, Ann. Phys. \bf 92\rm,
(1975), 142
\smallskip
\item{[\the\Banks]} T. Banks and E. Martinec, Nucl. Phys. \bf B249\rm,
(1987), 142
\smallskip
\item{[\the\deSan]} N.E.~Mavromatos, J.L.~Miramontes and
J.M.~S\'anchez de Santos, Phys. Rev. \bf D40\rm, (1989), 535
\smallskip
\item{[\the\pf]} B.P.~Dolan, {\it Integrability Conditions For
Potential Flow Of} \hfill\break
{\it The Renormalisation Group} to appear in Phys.~Lett.~B
\smallskip
\item{[\the\Metric]} S. Amari, \it Differential-Geometric Methods In Statistics
\hfill\break\rm Lecture Notes In Statistics, Vol 28, Springer-Verlag (1985)
\smallskip
\item{[\the\Denjoe]} D. O'Connor and C.R. Stephens,
\it Geometry The Renormalisation
Group And Gravity \rm\hfill\break
D.I.A.S. Preprint (1992)
\smallskip
\item{[\the\Zamo]} A.B Zamolodchikov, Rev. Math. Phys. \bf 1\rm, (1990),
197
\smallskip
\item{[\the\HughIan]} I.~Jack and H.~Osborn, Nucl.~Phys.~\bf B343\rm, (1990),
647
\smallskip
\item{[\the\Zinn]} E.~Br\'ezin, J.C.~le Guillou and J.~Zinn-Justin,
Phys. Rev. \bf 10\rm, (1974), 892
\smallskip

\vfill\eject\end